# Apple Machine Learning Algorithms Successfully Detect Colon Cancer but Fail to Predict KRAS Mutation Status

Andrew A. Borkowski, MD*[1,2], Catherine P. Wilson, MT[1], Steven A. Borkowski[1], L. Brannon Thomas, MD[1,2], Lauren A. Deland, RN[1], Stephen M. Mastorides, MD[1,2]

[1]Pathology and Laboratory Medicine Service, James A. Haley VA Hospital, Tampa, Florida, USA.

[2]Department of Pathology and Cell Biology, University of South Florida, Morsani College of Medicine, Tampa, Florida, USA.

*aborkows@health.usf.edu
Pathology and Laboratory Medicine Service (673/113)
James A Haley VA Hospital
13000 Bruce B. Downs Blvd
Tampa, FL 33612
Tel: 813-972-7525

**ABSTRACT**

Colon cancer is the second leading cause of cancer-related death in the United States of America. Its prognosis has significantly improved with the advancement of targeted therapies based on underlying molecular changes. The KRAS mutation is one of the most frequent molecular alterations seen in colon cancer and its presence can affect treatment selection. We attempted to use Apple machine learning algorithms to diagnose colon cancer and predict the KRAS mutation status from histopathological images. We captured 250 colon cancer images and 250 benign colon tissue images. Half of colon cancer images were captured from KRAS mutation-positive tumors and another half from KRAS mutation-negative tumors. Next, we created Image Classifier Model using Apple CreateML machine learning module. The trained and validated model was able to successfully differentiate between colon cancer and benign colon tissue images with 98 % recall and 98 % precision. However, our model failed to reliably identify KRAS mutations, with the highest realized accuracy of 66 %. Although not yet perfected, in the near future Apple CreateML modules can be used in diagnostic smartphone-based applications and potentially alleviate shortages of medical professionals in understaffed parts of the world.

1. INTRODUCTION

**Background and significance**

Colon cancer is the second leading cause of cancer-related death in the United States of America. According to cancer statistics provided by the American Cancer Society, it is estimated that during 2018, 140,250 Americans will be diagnosed with colon cancer and 50,630 will die from it.[1] Prognosis in colon cancer has improved with the advancement of targeted therapies based on underlying molecular changes.[2] One of the most frequent molecular changes in colon cancer is activation of the KRAS proto-oncogene.[3] KRAS mutation status is also the main predictive biomarker for treatment selection in colon cancer.[4], [5]

Recent research suggests that using artificial intelligence, one can successfully classify and make mutation predictions from non-small cell lung cancer histopathological images. It is reported that using deep learning technology, artificial intelligence (AI) is able to predict STK11, EGFR, FAT1, SETBP1, KRAS and TP53 gene mutations from pathology images with high accuracy.[6]

**Objective**

Our objective is to investigate the possibility of using Apple machine learning algorithms to diagnose colon cancer and assess the presence of the KRAS mutation from histopathological images.

2. MATERIALS AND METHODS

### 2.1 Image acquisition

Fifty cases of colon adenocarcinoma were retrieved from our molecular database; 25 contained KRAS mutations and 25 were negative for KRAS mutations. The 50 cases included 23 biopsies, 7 polypectomy specimens with invasive adenocarcinoma, and 20 resection specimens. 250 images of colon cancer (five from each case) were obtained using a Leica Microscope MC190 HD Camera (Leica Microsystems, Wetzlar, Germany) connected to an Olympus BX41 microscope (Olympus Corporation of the Americas, Center Valley, PA, USA) and the Leica Acquire 9072 software for Apple computers. All the images were captured at a resolution of 1024 x 768 pixels using a 60x dry objective and saved on a 2011 Apple iMac computer (Apple Inc, Cupertino, CA, USA) running macOS v10.13.6. In addition, 250 images of benign colonic mucosa were obtained from negative distal margin sections of resection specimens.

### 2.2 Xcode Image Classifier builder in Playground

As we described previously we used the Apple Xcode Playground v10 on a 2018 Apple MacBook Pro running macOS v10.14 to create ImageClassifier Model with following lines of code. [7]

```
import CreateMLUI
let builder = MLImageClassifierBuilder()
builder.showInLiveView()
```

We opened the assistant editor in Xcode and then ran the code. The live view displayed the image classifier UI. We then dragged the training folder for training the model and the testing folder to evaluate the model on the indicated locations in live view.

### 2.3 Experiment 1

In this experiment, we tested the Apple Create ML model to detect colon cancer histopathological images regardless of KRAS mutation status. We created two classes of images (250 images each) with the following labels: Benign (benign colon tissue) and AdenoCA (colon adenocarcinoma). 80% of the images were randomly assigned to the training folder and 20% to the testing folder. The training folder included two class labeled subfolders with training images (80% of total). The testing folder included two class labeled subfolders with testing images (20% of total). In our previous work, we found that the best results are obtained with 50 iterations and no additional augmentation of images.

We then trained and tested our model with 50 iterations, 10 times (Table 1).

### 2.4 Experiment 2

In this experiment, we tested the Apple Create ML model to differentiate between KRAS mutation-positive colon adenocarcinoma and KRAS mutation-negative colon adenocarcinoma histopathologic images. We created two classes of images (125 images each) with the following labels: KRAS+ (colon adenocarcinoma cases with KRAS mutation) and KRAS- (colon adenocarcinoma cases without KRAS mutation). The training folder included two class labeled subfolders with training images (80% of total). The testing subfolder included two class labeled subfolders with testing images (20% of total). We tested our model with 50 iterations, 10 times (Table 2).

Next, we tested the same image data sets with various image augmentations (crop, rotate, blur, expose, noise, flip, a combination of rotate and flip, a combination of rotate and noise) (Table 3).

### 3. RESULTS

When testing benign colon tissue images against colon adenocarcinoma images our model was able to differentiate between the two with 98% recall and 98% precision. In other words, our AI model could correctly differentiate benign colon tissue and malignant colon tissue 98% of the time.

When testing KRAS+ colon adenocarcinoma images against KRAS- colon adenocarcinoma images our model failed to adequately differentiate between the two. Although training accuracy was 100%, validation accuracy was 63% and testing accuracy was only 50% (same accuracy as flipping the coin).

Next, we tried to improve our model by adding various image augmentations. The best results were obtained with a rotation of images with testing accuracy increasing to 66%.

## 4. DISCUSSION

In this study, we designed several experiments to test the ability of an Apple Create ML model to distinguish the difference between KRAS mutation-positive colon adenocarcinomas and KRAS mutation-negative colon adenocarcinomas. The model accurately classified and distinguished between colon cancer versus benign colon images 98% of the time; however, the model had far less success distinguishing between a KRAS positive mutation image and KRAS negative mutation image.

The discovery of target therapies in cancer treatments, due to underlining molecular mechanisms, has led to a revolutionary wave of testing involving the identification of molecular mutations. [3] The identification of the KRAS mutation in colon adenocarcinoma indicates a major directional change in the patient's oncological treatment options. [2] Classification of molecular mutations such as KRAS is essential to the future of healthcare and cancer treatment directions.[4]

In the initial stages of AI program development, most applications were written and tested for large data groups. Programs were utilized in cloud environments with extremely large numbers of data sets.[8] Deep learning techniques were applied to many areas of medical research such as dermatology, pathology, and radiology.[9]–[15] Recent smaller scale incorporations of AI programs have shown even more promising results.[16] Smaller scale applications have been created for smartphones highlighting AI's potential for universal accessibility and ease of use.[17][18] Visual DX and other smartphone applications allow for a widespread image analysis with low cost and the ability to triage cases for further indicated medical treatments.[19] The widespread availability of handheld applications would alleviate some of the strain caused by global pathologist shortages.[20]

Recent studies with medical AI programs have demonstrated several advantages. AI programs have a potential for faster test completion and reported results, versus timing and labor involvement in other medical procedures.[21] Image analysis programs created for pathology may be able to conserve the amounts of tissue required to diagnose cancer from minimal tumor microenvironments.[21] Successful detection of underlying molecular mutations by use of AI image analysis may function to triage positive cancer images, lowering the cost of highly complex, time-consuming DNA detection assays.[6], [9]

## 5. CONCLUSIONS

Our model was able to successfully diagnose cancer but was unreliable in the detection of pathomorphological expressions of the KRAS mutation. Experiments with increased numbers of histopathological images may be able to distinguish underlying KRAS molecular mutations. Additional future experiments may include comparisons of the Apple CreateML model with Google cloud machine learning Cloud AutoML to see if timing or accuracy can be improved.[22], [23]

ACKNOWLEDGMENTS

None

FUNDING

This material is the result of work supported with resources and the use of facilities at the James A. Haley VA Hospital.

**TABLES**

| | Accuracy (%) | | | Recall (%) | | Precision (%) | |
|---|---|---|---|---|---|---|---|
| Run | Training | Validation | Testing | Benign | ColonCA | Benign | ColonCA |
| 1 | 100 | 100 | 98 | 98 | 98 | 98 | 98 |
| 2 | 100 | 100 | 98 | 98 | 98 | 98 | 98 |
| 3 | 100 | 100 | 98 | 98 | 98 | 98 | 98 |
| 4 | 100 | 100 | 99 | 98 | 100 | 100 | 98 |
| 5 | 100 | 100 | 98 | 98 | 98 | 98 | 98 |
| 6 | 100 | 100 | 98 | 98 | 98 | 98 | 98 |
| 7 | 100 | 100 | 98 | 98 | 98 | 98 | 98 |
| 8 | 100 | 100 | 98 | 98 | 98 | 98 | 98 |
| 9 | 100 | 100 | 98 | 98 | 98 | 98 | 98 |
| 10 | 100 | 100 | 98 | 98 | 98 | 98 | 98 |
| Median | 100 | 100 | 98 | 98 | 98 | 98 | 98 |
| Average | 100 | 100 | 98 | 98 | 98.2 | 98.2 | 98 |

Table 1. Training, validation, and testing. Colon cancer versus benign colon tissue images.

| | Accuracy % | | | Recall (%) | | Precision % | |
|---|---|---|---|---|---|---|---|
| Run | Training | Validation | Testing | ColonCA (KRAS+) | ColonCA (KRAS-) | ColonCA (KRAS+) | ColonCA (KRAS-) |
| 1 | 100 | 63 | 50 | 44 | 56 | 50 | 50 |
| 2 | 100 | 63 | 50 | 44 | 56 | 50 | 50 |
| 3 | 100 | 63 | 50 | 44 | 56 | 50 | 50 |
| 4 | 100 | 63 | 50 | 44 | 56 | 50 | 50 |
| 5 | 100 | 50 | 48 | 40 | 56 | 48 | 48 |
| 6 | 100 | 63 | 50 | 44 | 56 | 50 | 50 |
| 7 | 100 | 50 | 48 | 40 | 56 | 48 | 48 |
| 8 | 100 | 63 | 50 | 44 | 56 | 50 | 50 |
| 9 | 100 | 50 | 48 | 40 | 56 | 48 | 48 |
| 10 | 100 | 63 | 50 | 44 | 56 | 50 | 50 |
| Median | 100 | 63 | 50 | 44 | 56 | 50 | 50 |
| Average | 100 | 59.1 | 49.4 | 42.667 | 56 | 49.4 | 49.4 |

Table 2. Training, validation, and testing. KRAS positive versus KRAS negative colon cancer images.

| Augmentation | Accuracy % | | | Recall (%) | | Precision % | |
| --- | --- | --- | --- | --- | --- | --- | --- |
| | Training | Validation | Testing | ColonCA (KRAS+) | ColonCA (KRAS-) | ColonCA (KRAS+) | ColonCA (KRAS-) |
| Crop | 100 | 50 | 52 | 40 | 64 | 53 | 52 |
| Rotate | 100 | 63 | 66 | 60 | 72 | 68 | 64 |
| Blur | 100 | 63 | 52 | 48 | 56 | 52 | 52 |
| Expose | 100 | 63 | 44 | 48 | 40 | 44 | 43 |
| Noise | 100 | 63 | 54 | 44 | 64 | 55 | 53 |
| Flip | 100 | 50 | 56 | 60 | 52 | 56 | 57 |
| Rotate + Flip | 100 | 63 | 50 | 60 | 40 | 50 | 50 |
| Rotate + Noise | 100 | 50 | 52 | 40 | 64 | 53 | 52 |

Table 3. Training, validation, and testing. KRAS positive versus KRAS negative colon cancer images with different image augmentations.